\documentclass[twocolumn]{elsart}

\usepackage{graphicx,harvard}

\usepackage{amssymb}

\def\url#1{{\ttfamily\def\/{/\discretionary{}{}{}}#1}}

\newcommand\neii{[Ne\,{\sc ii}]}

\begin{document}

\begin{frontmatter}
\title{The Minispiral in the Galactic Center revisited}
\author[HD]{Bernd Vollmer\thanksref{MEU}}
\thanks[MEU]{current address: Observatoire de Paris, Meudon, France}
\author[HD,BN]{Wolfgang J. Duschl}
\address[HD]{Institut f\"ur Theoretische Astrophysik, Tiergartenstr.\ 15, 
D-69121 Heidelberg, Germany}
\address[BN]{Max-Planck-Institut f\"ur Radioastronomie, Auf dem H\"ugel 69, 
D-53121 Bonn, Germany}

\begin{abstract}
We present the results of a re-examination of a \neii\ line emission data cube 
($\lambda$12.8$\mu$m) and discuss the kinematic structure of the inner $\sim 3 
\times 4$\,pc of the Galaxy. The quality of \neii\ as a tracer of ionized gas 
is examined by comparing it to radio data. A three dimensional representation 
of the data cube allows us to disentangle features which are projected onto the 
same location on the sky. A model of gas streams in different planes is fitted 
to the data. We find that most of the material is located in a main plane which 
itself is defined by the inner edge of the Circum-Nuclear Disk in the Galactic 
Center. Finally, we present a possible three dimensional model of the gas 
streams. 
\end{abstract}


\end{frontmatter}

\section{Introduction}
The dynamic behavior of the gas in the inner  few pc of the Galaxy is still a 
matter of debate and speculation. There is an almost disklike structure of 
neutral gas located around the Galactic Center (GC) \cite{GGWal87,JGGal93} 
\cite{SGWal86}. This {\it Circum-Nuclear Disk\/} (CND) extends over a radial 
distance of more than 5\,pc. It has an inner edge at about 2\,pc from the GC. 
The predominantly molecular gas moves around the GC with  a speed of 
$\sim$100\,km\,s$^{-1}$ corresponding to a Keplarian velocity for an enclosed 
mass of several 10$^{6}$\,M$_{\odot}$. Within the inner 2\,pc ionized gas has 
been found \cite{LC83} which has the appearance of a spiral and was therefore 
named the {\it minispiral\/}. The region within the CND is often referred to as 
{\it Sgr A West\/}. It can be separated into at least four different components 
\cite{LAS91} which we will discuss below. The kinematics derived from the 
\neii\ data shows a possible close link between this ionized material and the 
molecular gas at the inner rim of the CND. Thus the outer part of the 
minispiral appears to be the ionized edge of the CND. Often the minispiral was 
thought to be not only of spiral shape in projection against the celestial  
sphere, but rather a true one-armed spiral. However, the kinematics derived 
from radio recombination line observations \cite{RG93} indicated several 
different structures instead of one single spiral. 

In this paper we re-examine the \neii\ data and the kinematic information 
contained in it. First, we introduce the dataset that we use (Sect.\ 
\ref{sec:data}), and evaluate its usability as a tracer of ionized gas by 
comparing it with radio recombination lines (Sect.\ \ref{sec:radio}). In Sect.\ 
\ref{sec:vis}, we describe our visualization technique which proved to be of 
paramount importance for our analysis. Section \ref{sec:morph} is devoted to 
a discussion of the morphological details in the image of the minispiral. The 
model that we derive from the data is presented in Sect.\ \ref{sec:model}, in 
the subsequent Section, we describe the three dimensional structure of the gas 
streams, and summarize our conclusions in Sect \ref{sec:concl}.
  
\section{The data\label{sec:data}}
\begin{figure}
\resizebox{\hsize}{!}{\includegraphics{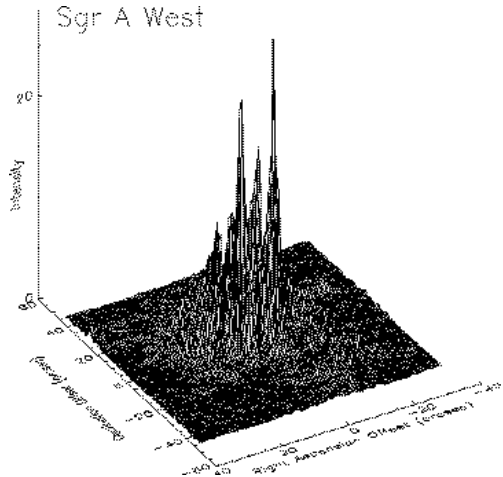}} \caption{Ruled-surface 
representation of the \neii\ line emission. The intensity is given in units of 
10$^{-2}$\,erg\,s$^{-1}$\,cm $^{-2}$\,sr$^{-1}$. All velocity channels are 
summed. \label{fig:map}}    
\end{figure}

We use the \neii ($\lambda 12.8\,\mu$m) line emission observations described in 
detail by \citeasnoun{LAS91}.  The data cube was made available to us by Lacy. 
The observations were carried out using a mid-infrared cryogenic echelle 
spectrograph. The spectral dispersion was about 16.5\,km\,s$^{-1}$ per pixel, 
the spectral resolution being about 2 pixels FWHM. An area of 
$75^{\prime\prime}$ $\times$ 90$^{\prime\prime}$ was mapped with 
2$^{\prime\prime}$ resolution. A \neii\ line map was produced by summing over 
the channels covering Doppler shifts from $+380$ to $-412$\,km\,s$^{-1}$. This 
map is shown as a ruled-surface representation in Figure \ref{fig:map}. 

\begin{figure}
\resizebox{\hsize}{!}{\includegraphics{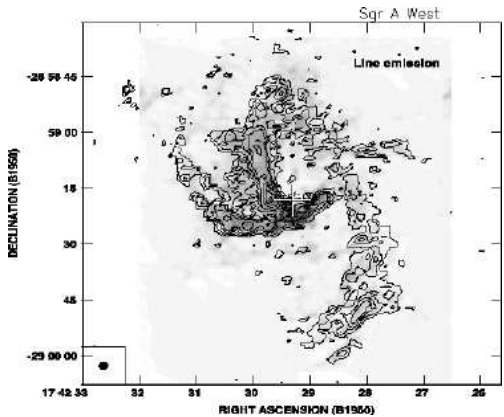}} \caption{Superposition of 
the 8.3\,GHz data as a contour representation together with the \neii\ line 
emission in a linear grey scale. Intensities greater than 
3\,10$^{-3}$\,erg\,s$^{-1}$\,cm$^{-2}$\,sr$^{-1}$ are set to this value in 
order to emphasize the features. The point source Sgr A* is indicated with a 
cross.\label{fig:radio}} 
\end{figure}

\section{The comparison with the radio data\label{sec:radio}}
\citeasnoun{RG93} observed the Sgr A West complex in the H92$\alpha$ line at 
8.3\,GHz with a resolution of 1$^{\prime\prime}$. As the data of the radio 
recombination line and the \neii\  emission at 12.8 $\mu$m give both 
informations about the distribution of the ionized gas, they should show the 
same features. Figure \ref{fig:radio} shows the superposition of these data. 
As, at this point, we are not interested in fine details but in much more in 
the overall distributions, the different resolutions of the two datasets pose 
no problem. We find that the overall features are the same in both wavelengths 
indicating that 
\begin{enumerate}
\item there are no depopulation effects affecting the \neii\ data,
\item \neii\ line emission is a good tracer for the ionized gas 
\end{enumerate}

\section{Three dimensional visualisation\label{sec:vis}}
\begin{figure*}
\resizebox{\hsize}{!}{\includegraphics{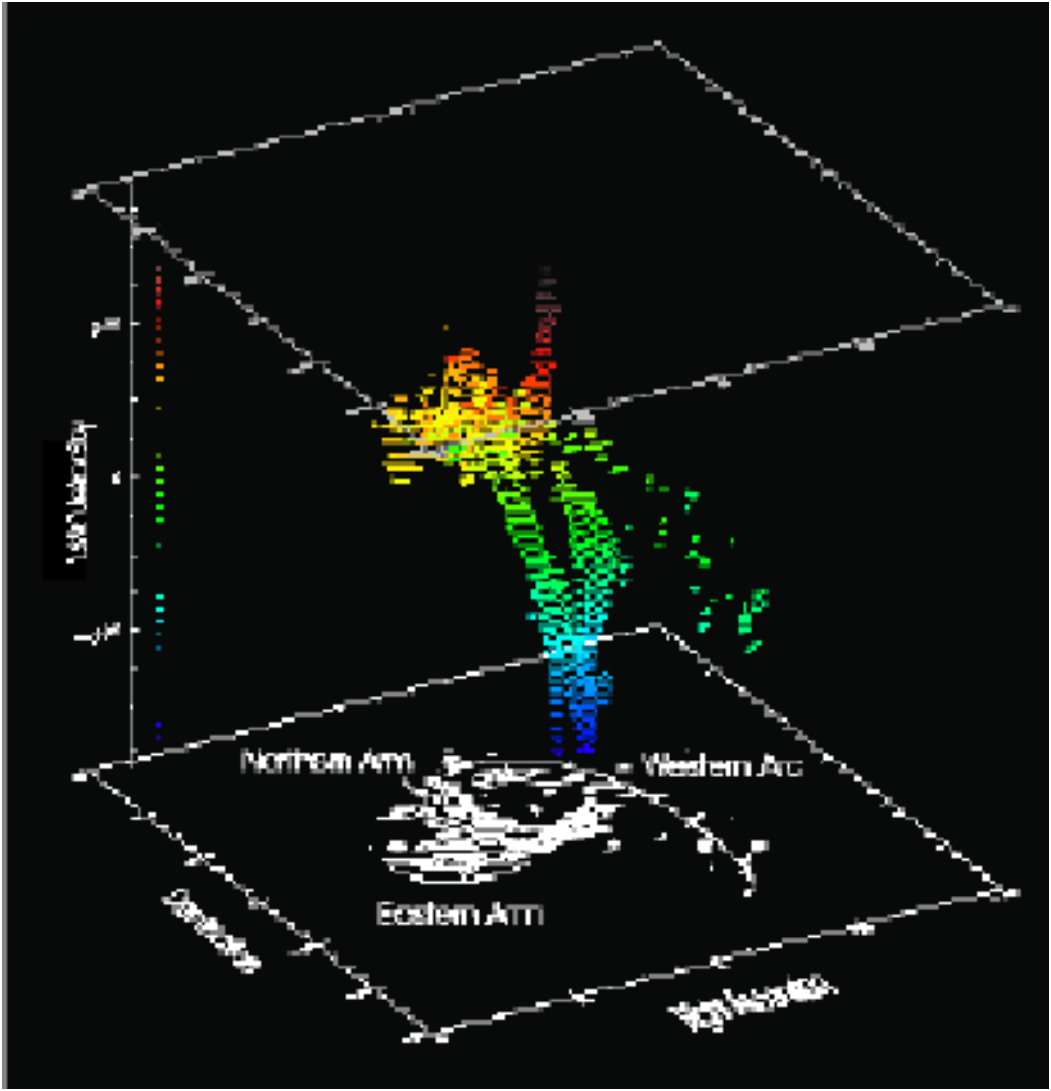}} 
\caption{A three dimensional representation of the data cube (radial velocity 
as a function of position on the sky. The velocities are color coded. In the 
bottom plane (R.A./Dec.), the projected minispiral is depicted. For an even 
clearer visualization of the 3D structure of the data cube, we also show it in 
an animation (\harvardurl{anim1.gif}).
\label{fig:color}} 
\end{figure*}

In the following, we will often use a three dimensional representation of the
data using a normal Cartesian coordinate system with R.A. corresponding to the 
$x$ direction, Dec. to the $y$ direction and the local standard of rest (LSR) 
velocities to the $z$ direction. To give a better visual impression, we have 
chosen to illuminate the shown surfaces from the observer's direction. In this 
representation features are the brighter the closer they are to the observer. 
The surfaces indicate pre-determined levels of constant intensity. To keep the 
figures as clear as possible, we give absolute numbers for the axes only in the 
color representation of Fig.\ \ref{fig:color}. 

\section{The morphology\label{sec:morph}}
\citeasnoun{LAS91} discussed four separate morphological components in the Sgr 
A West complex: 
\begin{enumerate}
\item the {\it Northern Arm\/} which is the most prominent feature and runs 
approximately from the radio source Sgr A* (the {\it Galactic Center\/}) to 
the north;
\item the {\it Western Arc\/} which is the bright, almost "vertical" rim in 
the west of Sgr A*;
\item the {\it Eastern Arm\/} which runs from the north east to the east of 
Sgr A*; and
\item the {\it Bar\/} which is situated "below" Sgr A* and and shows a straight 
shape.
\end{enumerate}

\begin{figure}
\resizebox{\hsize}{!}{\includegraphics{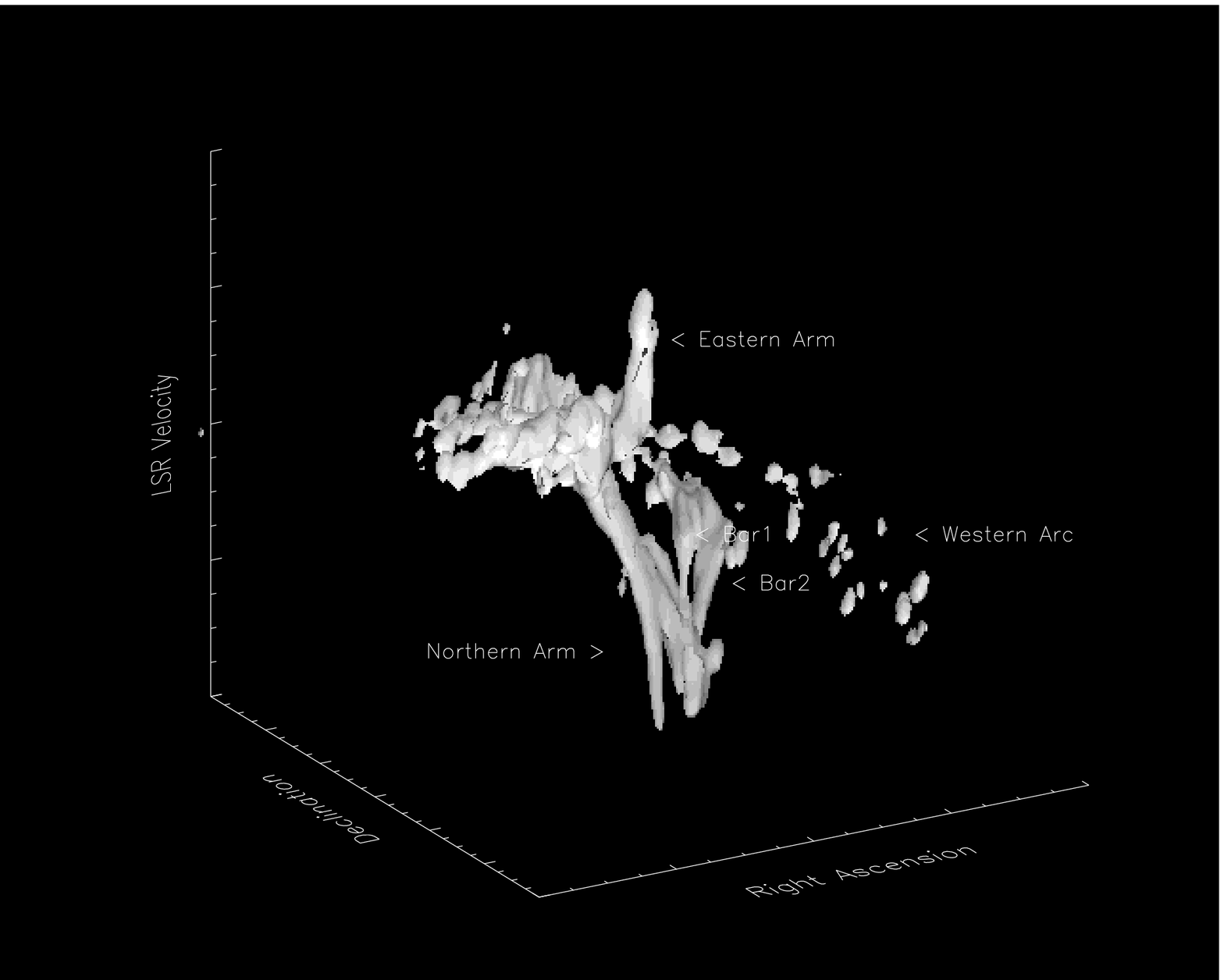}} \caption{The Sgr A West 
complex shown in an intensity level of 
1\,10$^{-2}$\,erg\,s$^{-1}$\,cm$^{-2}$\,sr$^{-1}$. For the absolute scaling, 
see Fig.\ \ref{fig:color}.\label{fig:three}} 
\end{figure}

\begin{figure}
\resizebox{\hsize}{!}{\includegraphics{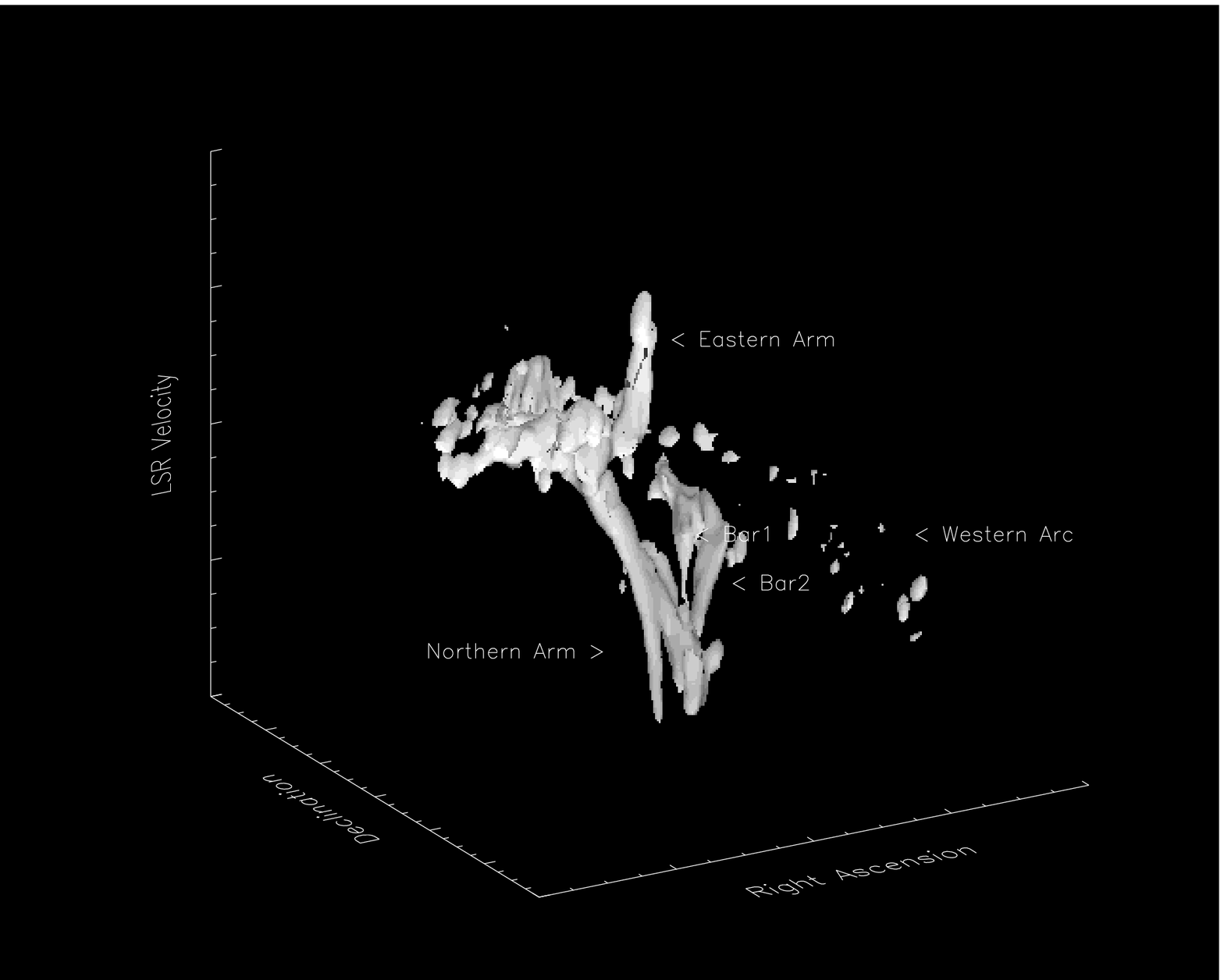}} \caption{The Sgr A West 
complex shown in an intensity level of 
6\,10$^{-3}$\,erg\,s$^{-1}$\,cm$^{-2}$\,sr$^{-1}$. For the absolute scaling, 
see Fig.\ \ref{fig:color}.\label{fig:four}} 
\end{figure}

\begin{figure}
\resizebox{\hsize}{!}{\includegraphics{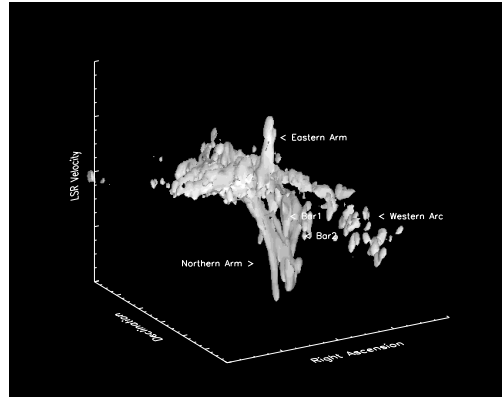}} \caption{The Sgr A West 
complex shown in an intensity level of 
3\,10$^{-3}$\,erg\,s$^{-1}$\,cm$^{-2}$\,sr$^{-1}$. This level is used for the 
comparison with the model calculations.For the absolute scaling, see Fig.\ 
\ref{fig:color}.\label{fig:five}} 
\end{figure}

\begin{figure}
\resizebox{\hsize}{!}{\includegraphics{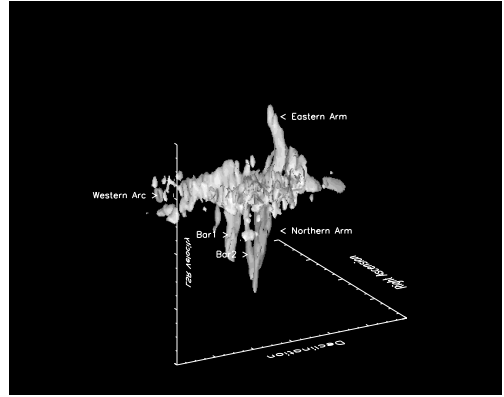}} \caption{The Sgr A West 
complex shown in an intensity level of 
3\,10$^{-3}$\,erg\,s$^{-1}$\,cm$^{-2}$\,sr$^{-1}$. The cube is rotated by 
120$^\circ$ with respect to the $z$-axis. Under this viewing angle, the two 
different components of the Bar can be seen particularly well.\label{fig:six}} 
\end{figure}

\begin{figure}
\resizebox{\hsize}{!}{\includegraphics{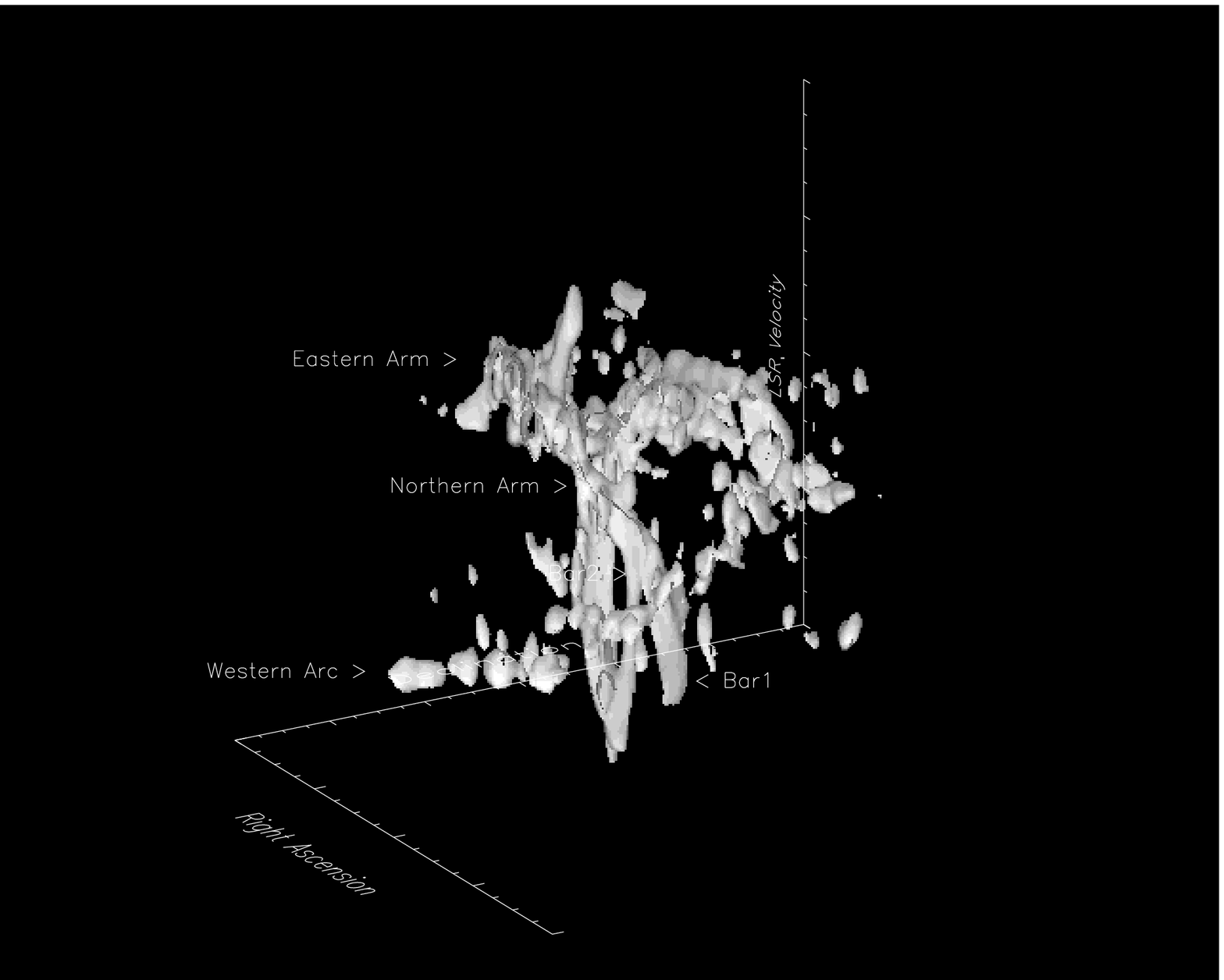}}
\caption{The same representation of the data cube as in Fig.\ \ref{fig:six}, 
but rotated by 300$^\circ$ with respect to the $z$-axis. Under this angle, the 
Western Arc can be seen particularly well as a bright line of emitting material 
in the foreground.\label{fig:seven}}
\end{figure}

Applying the above described analysis technique, it becomes evident that at 
least for the Bar the structure is more complicated than previoulsy thought.
Figures \ref{fig:three} to \ref{fig:five} show the data cube rotated by 
30$^\circ$ around the $x$- and $z$-axis for three different intensity levels. 
Figures \ref{fig:five}, \ref{fig:six}, and \ref{fig:seven} give different 
viewing angles to allow for a full appreciation of the three dimensional 
distribution.

The following features are marked:
\begin{enumerate}
\item {\bf The Northern Arm:} It appears as an almost vertical tube where 
material with the highest densities can be found. If one looks at Figure 
\ref{fig:four}, it becomes apparent, that part of the high negative velocity
end of this feature runs into the area of the Bar.
\item {\bf The Western Arc:} It represents the feature of lowest intensity, 
and as such appears only in the images of the data cube that trace the lowest
intensities. There it can be seen as a bent tube with a small velocity gradient.
\item {\bf The Eastern Arm:} It appears to consist of two components. A 
vertical {\it finger\/} of high intensity and a large {\it ribbon\/} which 
extends to the east of Sgr A*.
\item {\bf The Bar:} As the polarization measurements of \citeasnoun{AGSal91}
have already shown, this is the most complicated feature in the region. 
Looking at Figs. \ref{fig:three} and \ref{fig:four} it is clear that there are
two distinctly different components which we will call {\it Bar\,1\/} and {\it 
Bar\,2\/}.
\end{enumerate}

\section{The model\label{sec:model}}
For our model, we assume that the mass distribution in the inner few pc of 
the Galaxy can be reasonably well approximated by a spherically symmetric 
ansatz:
\begin{equation}
\frac{M(R)}{{\rm M}_\odot} = \frac{M_0}{{\rm M}_\odot} + 
1.6\,10^{6} \left(\frac{R}{\rm pc}\right)^{1.25}
\end{equation}
with $R$ the radius from the center of the Galaxy, $M(R)$ the radial mass 
distribution, and $M_0 = 3\,10^6\,{\rm M}_\odot$ the mass of the central black 
hole \cite{GEOE97}; the radially dependent contribution is an interpolation 
given by \citeasnoun{LDB93} for radii smaller than a few hundred pc. We assume 
the gaseous material to flow on closed circular orbits in the gravitational 
potential of this mass distribution. 

In order to allow for local turbulent motion (which however is assumed not to
dominate the flow patterns), an additional random, however on average isotropic 
velocity component is introduced. This corresponds -- through hydrostatic 
equilibrium -- to a certain thickness of the different flow patterns.

\begin{figure}
\resizebox{\hsize}{!}{\includegraphics{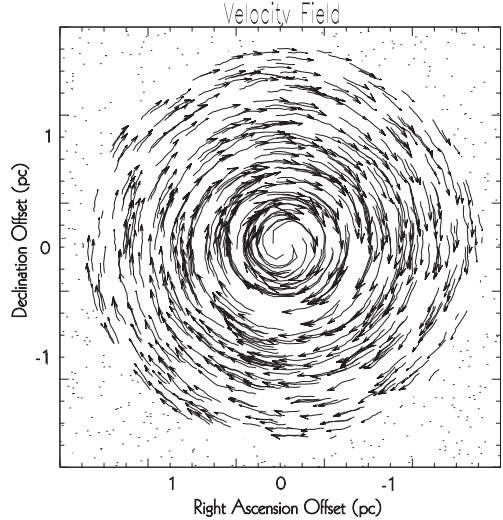}}
\caption{A representative velocity distribution for the disc model. The length 
of the arrows corresponds to the absolute value of the velocity.\label{fig:eight}}
\end{figure}

As, at this point we are not heading for a self-consistent hydrodynamic model 
of the flow, we finally allow for an ad hoc radial accretion velocity of 5\% of 
the local keplarian azimuthal velocity is assumed. We find that the model is 
much more sensitive to the absolute values of this radial velocity than of the 
turbulent velocity. However, given that the accretion velocity has a specified 
preferred direction (inwards) while the turbulent velocity is isotropic, this 
is not really a surprise. In Fig.\ \ref{fig:eight}, we show the projected 
velocity vectors for a disklike flow with a turbulent velocity of 40\%\ of the 
Keplerian velocity, and a turbulent velocity of 5\% of the Keplerian velocity. 

Choosing the turbulent velocity, i.e., the thickness of the disk, and the 
accretion velocity, in terms of an accretion disk model would be equivalent to 
prescribing the value of the viscosity. In terms of the standard theory of 
accretion disks \citeaffixed{FKR92}{see, e.g.,}, this corresponds to a 
viscosity parameter $\alpha \sim 0.3$, i.e., a value well within what one finds 
for other types of disks. 

We now model the flow by searching for the minimum number of different planes 
that are required to cover the observed features in the data cube. This 
obviously requires us not to think in terms of completely filled accretion 
disks. What we have in mind is a much more transient phenomenon in the 
following sense: We assume that clumps of gas -- undergoing some type of 
viscous interaction and/or tidal stretching -- are falling towards the GC. 
Locally they are described by a flow with Keplerian plus turbulent plus 
accretion motion. 

We find that the smallest number of planes to fit the whole measured data cube 
is three. One of them coincides with that of the CND. We will call that plane
the main plane, or plane (i). In Tab.\ \ref{tab:one} we give the orientations 
of the three planes. While for a single planes, some uncertainty about the sign 
of the inclination angle would be left, for the ensemble of three planes the 
consistency indicates the correct choice. 

\begin{table}
\caption{Orientation of the three planes.}
\label{tab:one}
\begin{tabular}{lrrc}
\hline
\noalign{\smallskip}
        &  position          &        &         \\
plane   &  angle             & incli- & closest \\ 
        &  (N$\rightarrow$E) & nation & edge    \\
\noalign{\smallskip} 
\hline 
\noalign{\smallskip} 
(i)   &  28$^\circ$ &  25$^\circ$ & W  \\ 
(ii)  & 132$^\circ$ & -15$^\circ$ & SW \\ 
(iii) & 115$^\circ$ &  20$^\circ$ & NE \\ 
\noalign{\smallskip} 
\hline 
\end{tabular}
\end{table}

\begin{figure}
\resizebox{\hsize}{!}{\includegraphics{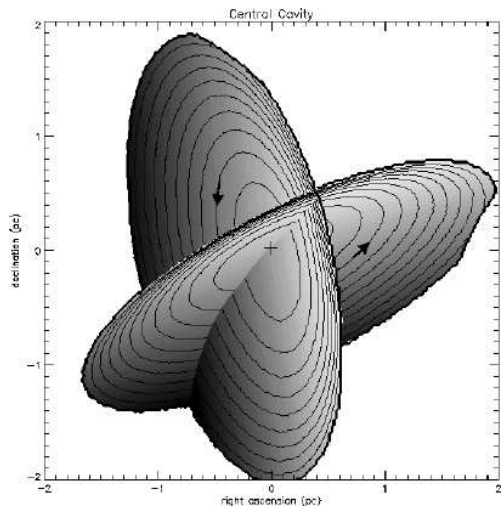}}
\caption{The relative orientation of planes (i) (top-bottom) and (iii). The 
arrows indicate the sense of rotation. The cross shows the position of 
SgrA*.\label{fig:nine}}
\end{figure}

To give an idea about the realtive orientations, in Fig.\ \ref{fig:nine} we 
show the orientation of planes (i) and (iii). The arrows indicate the senses of 
rotation. The cross marks the position of SgrA*. Plane (i) is the one that is 
oriented roughly top-bottom. In plane (i) the disk's velocity is positive for 
positive declination offsets and negative for negative declination offsets from 
Sgr A*. In the case of plane (iii) there are negative velocities for positive 
declination offsets and positive velocities for negative declination offsets. 
Thus one has a kind of counter-rotation of the material in plane (iii) with 
respect to that in plane (i).

\section{Comparing the model with the data}

\begin{figure}
\resizebox{\hsize}{!}{\includegraphics{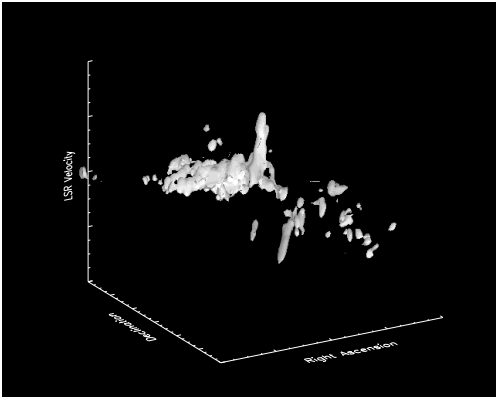}}
\caption{Residual of the data cube after subtracting plane (i) from 
it.\label{fig:ten}} 
\end{figure}

In order to evaluate the quality of the model fit, we determine the residuals 
of the subtration of the resulting model data cube from the observed one. At 
the same time, subtracting individual model planes from the entire observed 
data cube allows to demonstrate where individual structures are located.
As an example, in Fig.\ \ref{fig:ten} we show the data cube after subtraction 
of plane (i).

When performing this for all three planes, we find that
\begin{enumerate}
\item the Western Arc, the Northern Arm and the Bar\,1 are contained in
plane (i), whereas the whole Eastern Arm and the Bar\,2 are situated beyond;
\item plane (ii) covers the Finger of the Eastern Arm and the Bar\,2; 
\item the rest of the Eastern Arm is situated in plane (iii).
\end{enumerate}

\begin{figure}
\resizebox{\hsize}{!}{\includegraphics{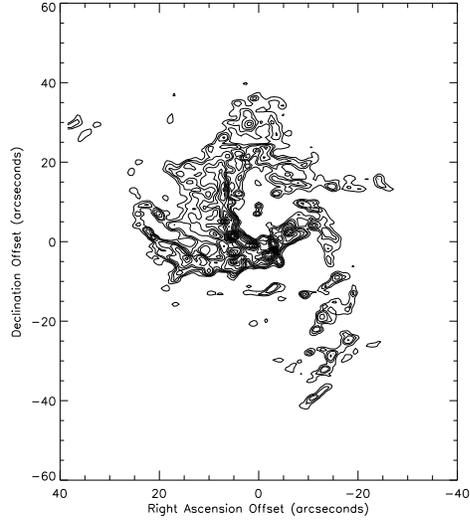}}
\\
\resizebox{\hsize}{!}{\includegraphics{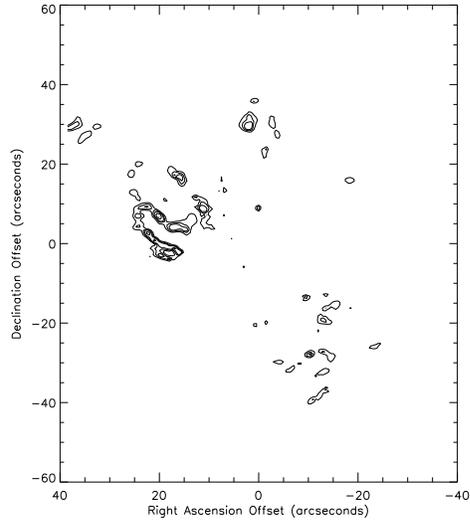}}
\caption{Contour representation of the original data (top panel) and the final 
residual (bottom panel) with the same contour levels.\label{fig:eleven}}
\end{figure}

In Fig.\ \ref{fig:eleven} we give the residual between observed and model data 
cube after subtracting all three planes and projecting it onto the sky. The 
only remaining structures are a small part of the Eastern Arm and features in 
the immediate vicinity of the IRS\,8 complex in the north. All other structures 
can be represented well within our model of the flow being mainly confined to 
three planes. 

This indicates that the Western Arc, the Northern Arm and the Bar\,1 are 
located in the main plane, i.e., the plane of the CND. What in projection 
appears a one structure called Eastern Arm seems to consist actually of at 
least two components, one of which lies in the same plane as the Bar\,2, while 
the remainder is located in a different plane.

\section{The distribution of the material in three spatial dimensions}
\begin{figure}
\resizebox{\hsize}{!}{\includegraphics{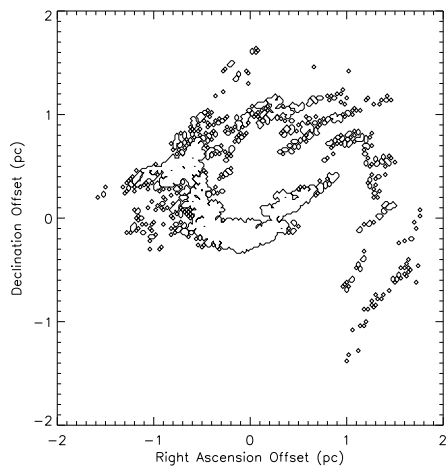}}
\caption{Contour plot of the distribution of emitting material in plane (i) seen 
face-onto that plane. The granular structure is an artefact due to discretization 
in the model.\label{fig:twelve}}
\end{figure}

\begin{figure}
\resizebox{\hsize}{!}{\includegraphics{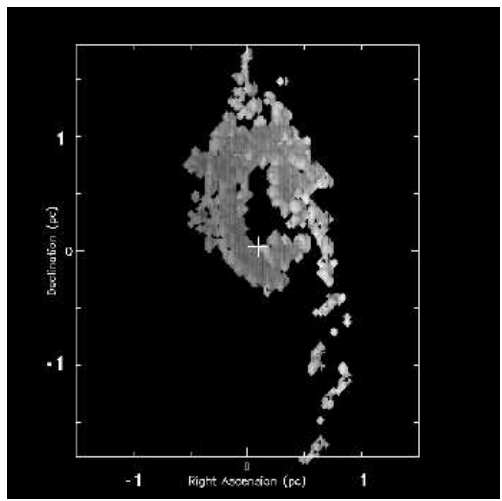}}
\caption{The distribution of Fig. \ref{fig:twelve} projected as an observer on 
Earth sees it.\label{fig:thirteen}}
\end{figure}

With the knowledge gained in the previous Section, one can now invert the 
problem and determine the flow patterns in the three planes, independent of the 
projected features. In Fig.\ \ref{fig:twelve} do so for plane (i), the main 
plane which coincides with the plane of the CND. In the Fig., we show a view 
face-onto plane (i). It shows the Northern Arm at the left of the center, the 
Bar\,1 at the right and the Western Arc which is bent from the end of the 
Northern Arm to a point below the center. Through projection an observer on 
Earth see the distribution shown in Fig.\ \ref{fig:thirteen}: One finds the 
shapes of the Northern Arm and the Western Arc. 

\begin{figure}
\resizebox{\hsize}{!}{\includegraphics{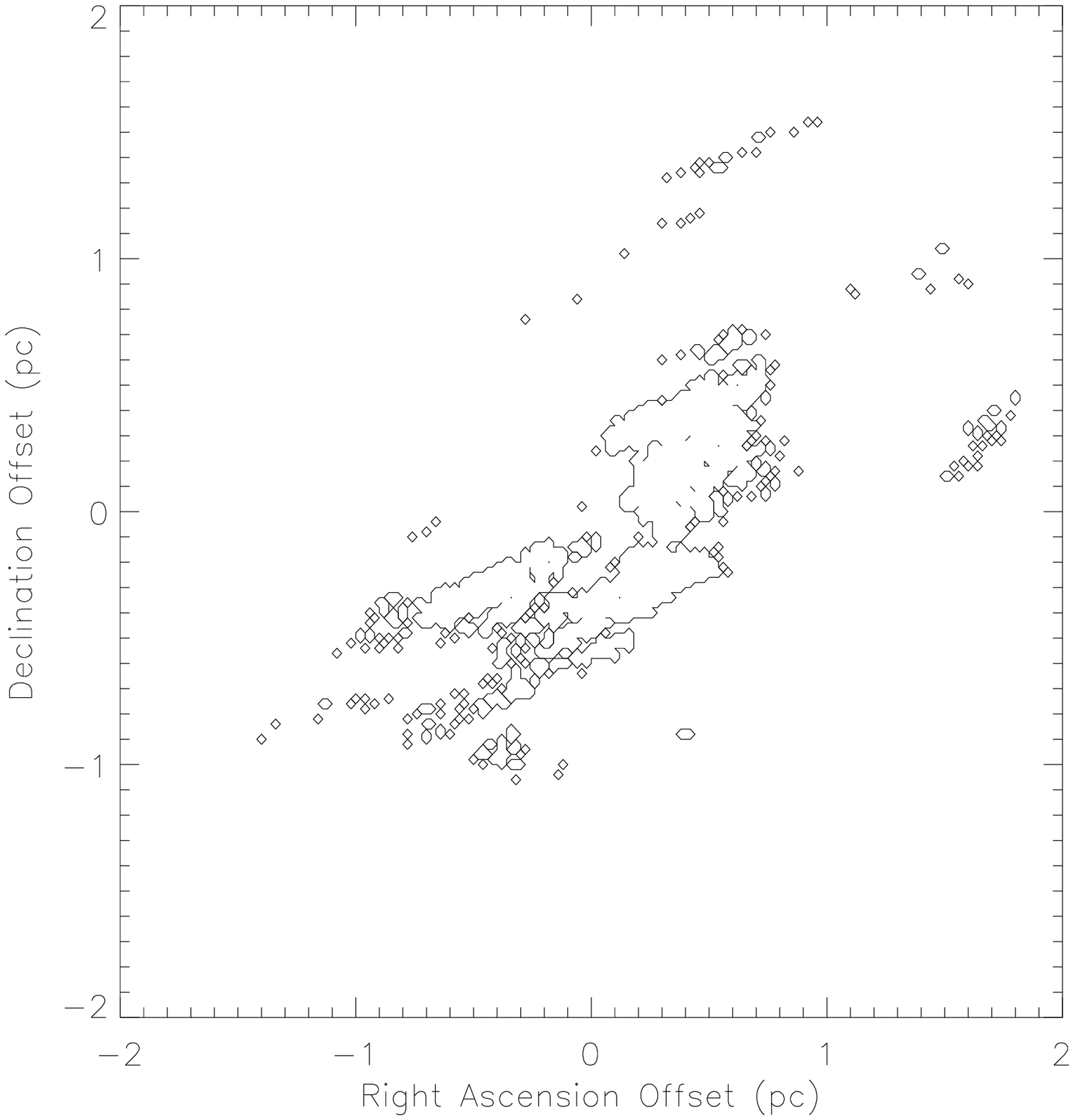}}
\\
\resizebox{\hsize}{!}{\includegraphics{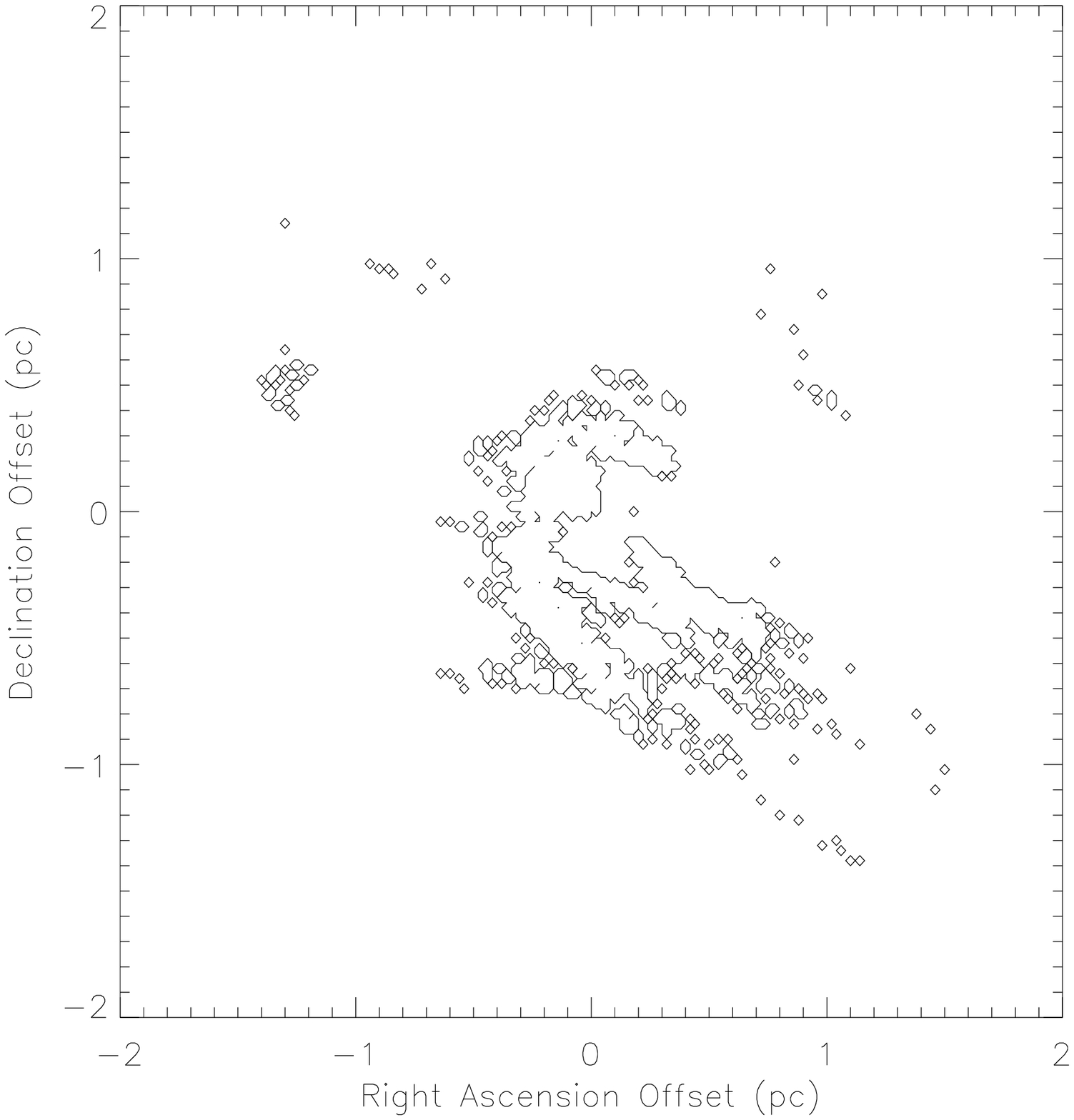}}
\caption{Contour representation the distribution of emitting material in planes 
(ii) and (iii). Both plans are shown face-on.\label{fig:fourteen}}
\end{figure}

The same procedures were repeated for planes (ii) and (iii). The results are 
shown in Figs.\ \ref{fig:fourteen}.

\begin{figure}
\resizebox{\hsize}{!}{\includegraphics{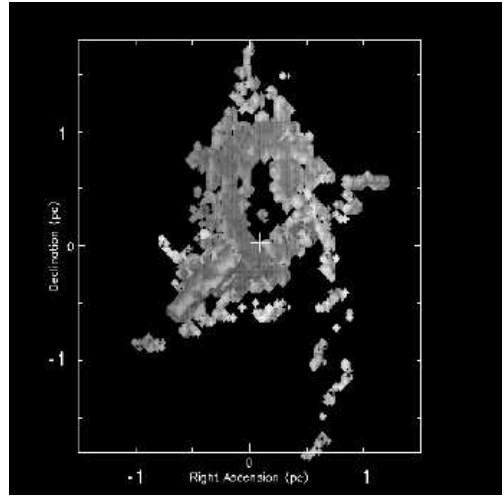}}
\caption{The ionized gas distribution of all three planes shown as an observer
on Earth sees it.\label{fig:fifteen}}
\end{figure}

Figure \ref{fig:fifteen} shows the combined projection of all three planes for 
an Earth-based observer providing a remarkably nice fit of the measured \neii\ 
data.
 
\begin{figure}
\resizebox{\hsize}{!}{\includegraphics{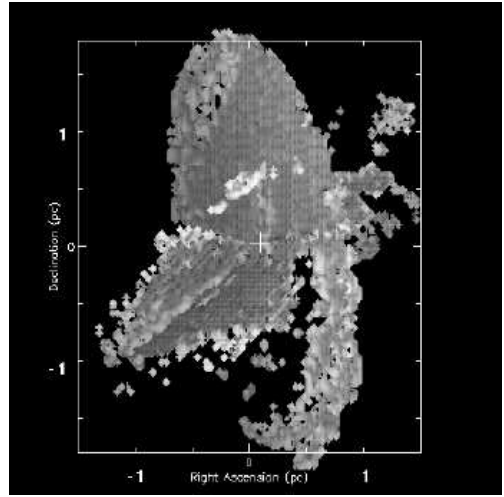}}
\caption{Same representation as in Fig. \ref{fig:fifteen}, however for a lower 
intensity level.\label{fig:sixteen}}
\end{figure}

Going to lower intensity levels, we find the possibility of a tenuous gas 
component which filling almost entirely plane (i). There are hints that this 
could also be the case for plane (ii), whereas the material in the third plain 
seems to form unconnected entities (Figure \ref{fig:sixteen}).

\begin{figure}
\resizebox{\hsize}{!}{\includegraphics{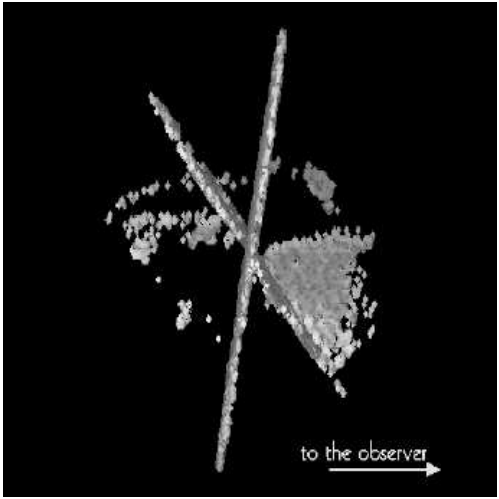}} \caption{The same data as 
in Figure \ref{fig:sixteen} rotated by an angle of 155$^\circ$ with respect to 
the $z$-axis. Plane (i) is seen vertically edge-on; plane (ii) is also seen 
edge-on. For an even clearer visualization of the 3D structure of the data 
cube, we also show it in an animation (\harvardurl{anim2.gif}). 
\label{fig:seventeen}} 
\end{figure}

Finally, we give in Fig.\ \ref{fig:seventeen} a view of the same data, however 
seen from a different angle (155$^\circ$ with respect to the $z$-axis) chosen 
such that planes (i) and (ii) are seen edge-on. 

\section{Conclusion\label{sec:concl}}
We have re-analyzed a \neii\ emission line distribution in the immediate 
vicinity of the GC by means of a three dimensional visualization of the data 
cube. We find that due to projection effects some physically distinct entities 
mislead us by appearing as single features. Based on the data cube, we have 
re-classified the structures in Sgr A West. We find that they are confined to 
mainly three distinct planes. Within the planes they represent the denser 
material. However, at least in two of the three planes, tenuous gas seems to 
fill almost the entire planes, not unlike accretion disks. Most of the 
minispiral's material is located in a main plane coinciding with that of the 
CND. 

We find the best fit for material moving in the planes on Keplerian circular 
orbits, overlayed with a turbulent velocity component of some 40\% of the 
Keplerian speed and an inwards radial velocity of 5\% of the orital velocity.

It is important to note that our fits exclude predominant outwards motion; it 
seems that a scenario of a mass stream from outside towards a central black 
hole is supported by our results. It is tempting to speculate that the 
different components that project as the minispiral onto the sky actually are 
chunks of matter ''falling out of the CND towards the black hole" (see also 
\citeasnoun{ZG99}). Then the presence of individual planes would be of little 
surprise. They are -- more or less randomly -- defined by the local angular 
momentum of the material leaving the CND. One expects -- on average -- a 
direction coinciding with that of the CND. However, as individual clumps may 
very well have an angular momentum that differs somewhat from the local 
average, for individual streams, planes different from that of the CND are 
possible. At the same time this tells us that these features are rather 
short-lived. 

The three dimensional spatial reconstruction of the distribution of the matter 
moreover indicates that counterrotating features are present. At this point it 
is suggestive but not yet clear whether this has anything to do with the 
counter-rotating early type stars found by \citeasnoun{GTKal96}. One can -- as 
a speculation -- not even exclude that the material in plane (iii) is indeed 
the rest of the material out of which those stars were formed. 

\section*{Acknowledgements}
We thank J.H.\ Lacy for making his data available to us, and D.A.\ Roberts and 
W.M.\ Goss for allowing is to use their radio image, in this paper Figure 2. We 
benefitted much from discussion of the topic of this paper with P.G.\ Mezger. 
The help of M.-F.\ Landr\'ea in the preparation of the animations is highly 
acknowledged.

\end{document}